\documentclass[prb,twocolumn,showpacs,preprintnumbers,amsmath,amssymb,superscriptaddress]{revtex4}

\usepackage{graphicx}
\usepackage{dcolumn}
\usepackage{bm}

\begin{document}


\title{Magnetic field induced quantum phase transition of the 
$\bm{S = 1/2}$ antiferromagnet K$_2$NaCrO$_8$} 

\author{S.~Nellutla}
\altaffiliation[Present address: ]{Department of Chemistry, 
North Carolina State University, Box 8204, Raleigh, NC 27695}
\affiliation{National High Magnetic Field Laboratory, Tallahassee, FL 32310}
\author{M.~Pati}
\affiliation{Department of Chemistry and Biochemistry, Florida State 
University, Tallahassee, FL 32306}
\author{Y.-J.~Jo}
\author{H.~D.~Zhou}
\affiliation{National High Magnetic Field Laboratory, Tallahassee, FL 32310}
\author{B.~H.~Moon}
\author{D.~M.~Pajerowski}
\affiliation{Department of Physics, University of Florida, Gainesville, FL 
32611-8440}
\affiliation{National High Magnetic Field Laboratory, Gainesville, FL 32611-8440}
\author{Y.~Yoshida}
\altaffiliation[Present address: ]{Institute of Applied Physics and 
Microstructure Research Center, University of Hamburg, D-20355 Hamburg, 
Germany}
\affiliation{Department of Physics, University of Florida, Gainesville, FL 
32611-8440}
\affiliation{National High Magnetic Field Laboratory, Gainesville, FL 32611-8440}
\author{J.~A.~Janik}
\affiliation{National High Magnetic Field Laboratory, Tallahassee, FL 32310}
\affiliation{Department of Physics, Florida State University, Tallahassee, FL 
32310}
\author{L.~Balicas}
\affiliation{National High Magnetic Field Laboratory, Tallahassee, FL 32310}
\author{Y.~Lee}
\author{M.~W.~Meisel}
\author{Y.~Takano}
\affiliation{Department of Physics, University of Florida, Gainesville, FL 
32611-8440}
\affiliation{National High Magnetic Field Laboratory, Gainesville, FL 32611-8440}
\author{C.~R.~Wiebe}
\altaffiliation[Present address: ]{Department of Chemistry, University of Winnipeg, 
Winnipeg, MB R3B 2E9, Canada}
\affiliation{National High Magnetic Field Laboratory, Tallahassee, FL 32310}
\affiliation{Department of Physics, Florida State University, Tallahassee, FL 
32310}
\author{N.~S.~Dalal}
\affiliation{National High Magnetic Field Laboratory, Tallahassee, FL 32310}
\affiliation{Department of Chemistry and Biochemistry, Florida State 
University, Tallahassee, FL 32306}

\date{\today}

\begin{abstract}
The magnetic properties of alkali-metal peroxychromate K$_2$NaCrO$_8$ are governed 
by the $S = 1/2$ pentavalent chromium cation, Cr$^{5+}$.  
Specific heat, magnetocalorimetry, ac magnetic susceptibility, 
torque magnetometry, and inelastic neutron scattering data have been acquired over a 
wide range of temperature, down to 60~mK, and magnetic field, up to 18~T.  
The magnetic interactions are quasi-two-dimensional prior to long-range ordering,  
where $T_N = 1.66$~K in $H = 0$.  
In the $T \rightarrow 0$ limit, the 
magnetic field tuned antiferromagnetic-ferromagnetic phase transition suggests 
a critical field $H_c = 7.270$~T and a critical exponent $\alpha = 0.481 \pm 0.004$.  
The neutron data indicate the magnetic 
interactions may extend over intra-planar nearest-neighbors and inter-planar 
next-nearest-neighbor spins.
\end{abstract}

\pacs{75.40.Cx, 75.30.Kz, 75.50.Ee}
\maketitle

The study of the similarities and differences between classical and 
quantum phase transitions has a rich history, and recent research 
has focused on systems in reduced dimensions.\cite{Giamarchi1,Sachdev1,Sachdev2,Giamarchi2}   
For magnetic systems, the model Hamiltonian is often written as
\begin{equation}
\mathcal{H}\;=\;\sum_{(i,j)} \;J_{ij}\;\vec{S}_i \cdot \vec{S}_j\;+
\;\vec{H} \,\cdot \,\overleftrightarrow{g} \,\cdot\,\sum_i\,\vec{S}_i \;\;\;,
\end{equation}
where the first sum over $(i,j)$ is often restricted to nearest-neighbors $(nn)$ 
but may sometimes extend to include next-nearest neighbors $(nnn)$,  
$J>0$ indicates antiferromagnetic interactions, and the Zeeman (second) term 
includes the presence of the externally applied magnetic field $H$ and an 
anisotropic Land\'{e} $g$-tensor.  It is important to emphasize that both 
the spin and spatial dimensions are crucial elements of any model, and in 
this context, the critical exponents can be considered fingerprints 
for classifying the nature of the transition.  In particular, the 
order parameter describing the transition can be characterized by  
$(T_c - T)^{\beta}$ as the temperature $T$ approaches the 
critical temperature $T_c$ or, in the $T \rightarrow 0$ limit, by 
$(H_c - H)^{\alpha}$ for a quantum phase transition tuned by the magnetic field.
\cite{Giamarchi1,Sachdev1,Sachdev2,Giamarchi2,Troyer,Wessel,Nohadani,Sebastian1,Radu1,Sebastian2,Radu2}
\begin{figure}[ht]
\includegraphics[width=3.375in]{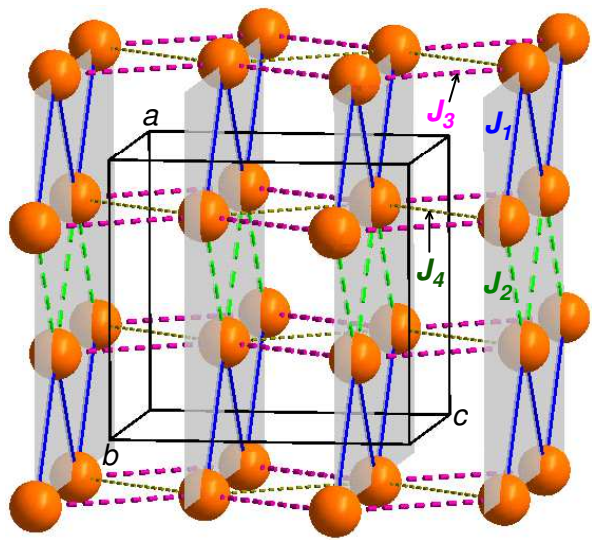}
\caption{(Color online) Spin network in K$_2$NaCrO$_8$ as seen along the $b$-axis, where 
only the Cr$^{5+}$ ions are shown for clarity.   
The vertical planes are the mirror planes containing $J_1$ (solid lines) and the 
$J_2$ (dashed lines) paths connecting the $nn$ Cr$^{5+}$ ions. Exchange paths corresponding 
to $nnn$ interactions are labeled as $J_3$ (short-dashed lines) and $J_4$ (dotted lines).}
\end{figure}

The purpose of the present work was to consider the curious case of 
the pentavalent chromium cation, Cr$^{5+}$, an $S = 1/2$ ($3d^1$) ion that is 
not a common choice for condensed matter 
research.\cite{Riesenfeld,Dickman,Cage,Jimenez,Nakajima,Singh,Kofu1,Choi,Kofu2}  
Specifically, the alkali-metal peroxychromate K$_2$NaCrO$_8$ crystallizes with  
orthorhombic symmetry (space group $Pbcm$) with the lattice parameters 
$a = 8.5883$~\AA, $b = 7.9825$~\AA, and $c = 9.2432$~\AA~at 173~K.\cite{Cage} 
The Cr$^{5+}$ ion is bonded 
to four peroxy (O$_2$)$^{2-}$ ligands in a dodecahedral geometry, and 
the ions form the three dimensional network shown in Fig.~1.\cite{Cage,Choi}   
The data from EPR studies showed the $g$-tensor is anisotropic, 
with $g_z = 1.9851$, $g_y = 1.9696$, and $g_x = 1.9636$, the electronic ground 
state is $d_{z^2}$, and the magnetic $z$-axis is along the [110] crystal direction.   
During studies of the static magnetic susceptibility, 
$\chi_{\mathrm{dc}}(H = 0.3$~T$,1.8$~K $\leq T \leq 20$~K), and 
the specific heat, $C(0 \leq H \leq 9$~T, $1.8$~K $\leq T \leq 10$~K), 
Cage and Dalal\cite{Cage} observed ordering signatures in polycrystalline samples. 
More specifically, while a rounded maximum was observed in 
$\chi_{\mathrm{dc}}(H = 0.3$~T, $T \approx 2.29$~K),
the $C(H \approx 0, T \approx 1.8$~K) data appeared to possess a $\lambda$-like peak. 
These signatures indicate the presence of low dimensional antiferromagnetic behavior 
and the onset of long-range antiferromagnetic ordering.  
More recently, Choi \emph{et al.}\cite{Choi} performed low temperature magnetization and 
$^{23}$Na NMR measurements and identified a quantum phase transition 
from an antiferromagnetically ordered state to a ferromagnetically polarized state at 
critical field $H_c$ of about 7.4~T, with $\beta = 0.44 \pm 0.05$ in $H = 5$~T and 
$\beta = 0.53 \pm 0.05$ in $H = 6.8$~T.  Since mean-field theory yields 
$\beta = 1/2$ for magnetization, these workers concluded that 
the high field magnetization of K$_2$NaCrO$_8$ is mean-field-like.

Herein, the results of specific heat, magnetocaloric effect, ac magnetic susceptibility, 
and torque magnetometry are reported to construct the magnetic phase diagram 
of K$_2$NaCrO$_8$, with particular emphasis on the $T \rightarrow 0$ limit of the magnetic 
field induced quantum phase transition. The data indicate the magnetic interactions 
are quasi-two-dimensional prior to long-range ordering.    
Inelastic neutron scattering studies in zero magnetic field confirm the antiferromagnetic 
transition occurs without any low temperature structural transitions and suggest 
the strongest magnetic interactions may include both $nn$ and $nnn$ ions. Furthermore, 
in the critical regime, the specific heat and magnetocaloric effect systematically identify 
the phase transition, while the magnetic measurements, although consistent with each other,
yield a slightly shifted trend as $T\rightarrow 0$.  

\begin{figure}[ht]
\includegraphics[width=3.36in]{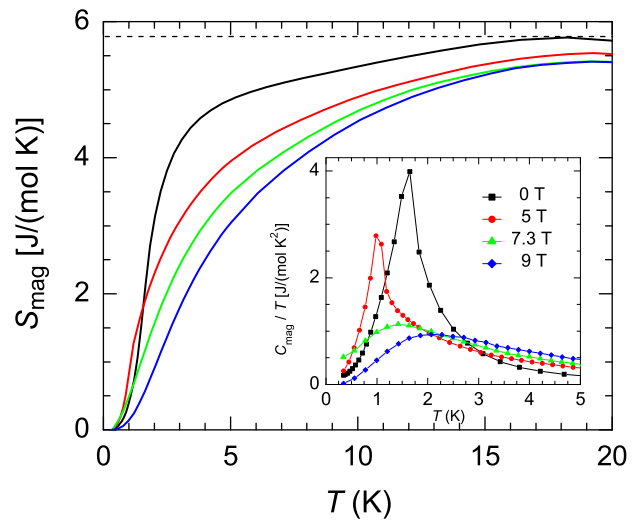}
\caption{(Color online)  The temperature dependence of the magnetic entropy, 
$S_{\mathrm{mag}}(T)$, at several representative magnetic fields.  The 
inset shows the temperature dependence of the magnetic specific heat divided 
by temperature, $C_{\mathrm{mag}}(T)/T$, at the same magnetic fields.   
$C_{\mathrm{mag}}(T)$ was obtained by subtracting a lattice contribution from the 
total specific heat, see text. $S_{\mathrm{mag}}(T)$ was obtained by integrating 
the data shown in the inset.   
As $T$ increases, $S_{\mathrm{mag}}(T)$ slowly approaches the $R \ln 2$ value, 
designated by the dashed line, 
as expected for a $S = 1/2$ system.  In $H=0$, about 60\% of the entropy is 
acquired above $T_N$, indicating significant contributions from short range correlations.}
\end{figure} 

Single crystals of K$_2$NaCrO$_8$ were prepared by modification of the method of 
Riesenfeld.\cite{Riesenfeld} The synthesis involves the reduction of Cr$^{6+}$ by 30\% 
H$_2$O$_2$ at 5$^{\circ}$C in the 1:1 molar mixture of aqueous NaOH and KOH.  
Specific heat and magnetocaloric data were collected on single crystals with $H$ applied 
along the $c$-axis. The specific heat studies employed the relaxation technique, 
150~mK $\leq T \leq$ 20~K, with $H \leq 18$~T using two different instruments.
The magnetic specific heat, $C_{\mathrm{mag}}$, was obtained by subtracting the zero-field 
Debye contribution, namely $C_{\mathrm{phonon}} = (1.85 \times 10^{-3})\,T^3$ J/(mol K$^4$), 
from the total specific heat at all fields, and was integrated to yield the magnetic entropy, 
$S_{\mathrm{mag}}(T)$, Fig.~2. 
The magnetocaloric data, 150~mK $\leq T \leq$ 600~mK, were collected in $H \leq 18$~T 
with a field sweep rate of 0.2~T/min, Fig.~3a.  
For the ac susceptibility, $\chi_{\mathrm{ac}}$, a sample consisting of a random 
arrangement of microcrystallites (8.44~mg) was mounted, using GE varnish dissolved in ethanol, 
on a silver strip and wrapped in Kapton tape.  The measurements were performed in a homemade 
dilution refrigerator probe using a standard set of inductively coupled coils 
operating at 1.114~kHz, Fig.~3b.  
Torque magnetometry, performed down to 280~mK and up to 18~T, utilized a single crystal sample 
(1.2~mg) mounted using Stycast epoxy on a 0.005~inch thick CuBe cantilever soldered to a post. 
The torque experienced by the sample deflects the cantilever,  
and the deflection was measured capacitively through a manual capacitance bridge excited at 
30~V and 5~kHz, Fig.~3c.  
Zero field neutron scattering work down to 1.4~K used a powder-like sample, generated by 
grinding microcrystals, that was mounted on the NIST-CNS disk chopper spectrometer (DCS) 
tuned to a wavelength of 4.8~\AA.  

\begin{figure}[ht]
\includegraphics[width=3.375in]{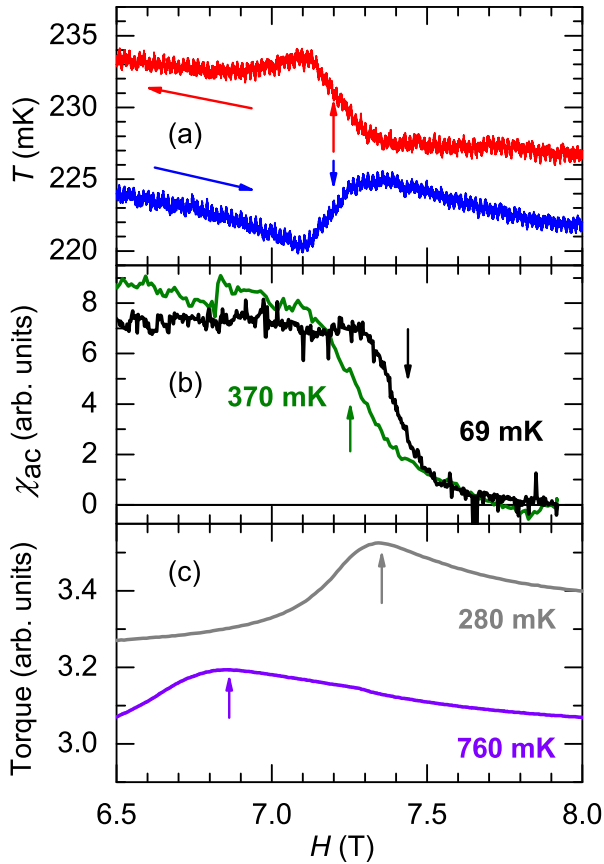}
\caption{(Color online) 
(a) The magnetocaloric effect curves for up and down field sweeps near 230~mK.  
(b) The ac susceptibility, $\chi_{\mathrm{ac}}(H)$ at 69~mK and 370~mK. 
(c) The isothermal magnetic torque measured as a function of $H$.  
In each panel, the arrows designate the positions taken as the field induced phase transition.}
\end{figure}

The distinct $\lambda$ anomaly observed in the magnetic specific heat (Fig.~2) 
at $T_N = 1.66$~K and $H = 0$ indicates that K$_2$NaCrO$_8$ undergoes 
long-range antiferromagnetic ordering (LRAFO) as conjectured from earlier specific heat 
measurements.\cite{Cage} The LRAFO is suppressed in the presence of a magnetic field, as
$T_N$ shifts slowly to lower temperatures until $H$ reaches about 5~T and 
rapidly thereafter until the LRAFO is completely suppressed around 7.3~T.  
For $H > 5$~T, the  Schottky-type anomaly overlapping 
the LRAFO at zero field becomes the dominant feature of $C_{\mathrm{mag}}(T)$, 
and the corresponding position in temperature of the heat capacity maximum shifts to 
higher temperatures.  Consequently, the position of the phase transition is harder 
to identify.  On the other hand, the magnetocaloric effect produces peaks in the up 
and down sweep directions, and these features correspond to the magnetic phase transition, 
while the center of these anomalies is identified as the transition point, Fig.~3.  
In Fig.~4, the transition points identified by these two methods are shown to yield 
a single trend of the phase boundary between the LRAFO and the polarized 
ferromagnetic state.  

\begin{figure}[ht]
\includegraphics[width=3.375in]{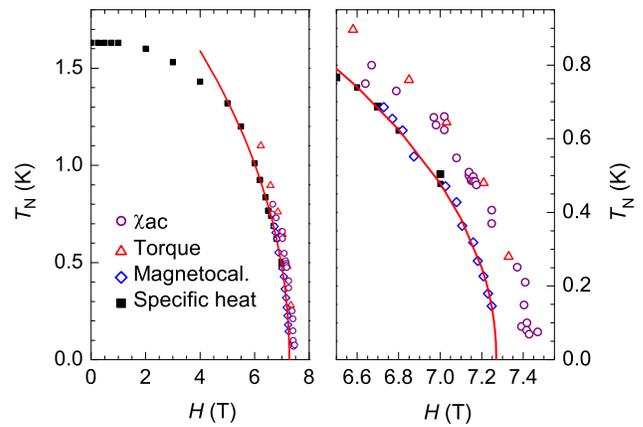}
\caption{(Color online) Magnetic phase diagram of K$_2$NaCrO$_8$. 
The solid line corresponds to Eq.~2 with $H_c = 7.27$~T and $\alpha = 0.481 \pm 0.004$.  
The right panel provides an expanded view near $H_c(T \rightarrow 0)$, where the 
specific heat and magnetocaloric transition points are 
systematically different than the ones extracted from standard analysis of the 
magnetic measurements, $\chi_{\mathrm{ac}}$ and torque.}
\end{figure}
\begin{figure}[ht]
\includegraphics[width=3.375in]{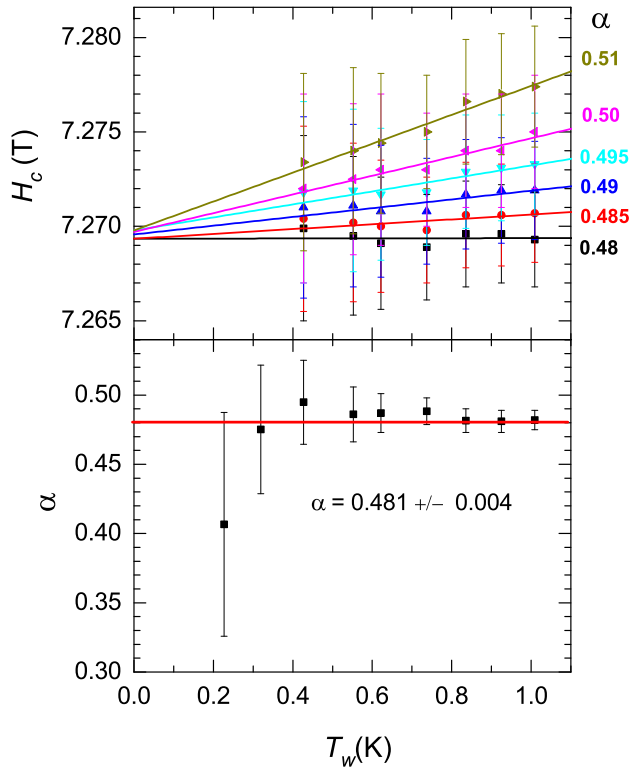}
\caption{(Color online) Estimates of (top panel) the critical field $H_c$ and 
(bottom panel) the critical exponent $\alpha$ obtained from the ``$T$-window'' analysis 
of the transition, see text for detailed description.  
The results are $H_c = 7.2696 \pm 0.0002$~T and $\alpha = 0.481 \pm 0.004$.}
\end{figure}

With respect to the magnetic studies using $\chi_{\mathrm{ac}}$ and torque, the 
signatures designating the critical field of the transition are sharpest as 
$T \rightarrow 0$.  For $\chi_{\mathrm{ac}}$, the inflection point between the 
low and high field regions is often used, Fig.~3b.  Alternatively, these data 
are integrated to yield the isothermal field dependence of the magnetization, 
$M(H)$, but an extrapolation protocol must be used as the 
location of the critical field is often smeared.\cite{Goddard,Xiao,Orendacova}
In our case, both routines 
provide essentially the same results that are plotted in Fig.~4.  For the torque 
measurement, the peak of the response appears on a smooth, nearly quadratic background, 
Fig.~3c, and is taken as the location of the transition, Fig.~4.  
It is interesting to note that both techniques indicate a 
phase boundary that is slightly shifted from the one given by the specific heat 
and magnetocaloric effect results, Fig.~4. 

\vskip 5mm

Since the response of the specific heat unambiguously marks the transition and 
the magnetocaloric effect data identify the same critical field, these results were 
used to fit to
\begin{equation}
T_N\;=\;A\,(H_c - H)^{\alpha}\;\;\;\;\;.
\end{equation}
Care must be used when fitting Eq.~2, as both $H_c$ and $\alpha$ must be 
resolved.\cite{Nohadani,Sebastian1,Radu1,Sebastian2,Radu2}  
Consequently, the ``$T$-window'' analysis and empirical convergence 
methods\cite{Sebastian1} were employed, by systematically varying the 
maximum temperature of the fits from 1~K to 230~mK, 
to extract the values of $H_c = 7.27$~T and $\alpha = 0.481 \pm 0.004$.  
In detail, to extract $H_c$, the low temperature ($T < 1$~K) phase boundary was 
fit to Eq.~2 for various $\alpha$ values in a temperature window, $T_W$, 
of decreasing size. For example, the phase boundary with a window size of 20 data points 
was fit to Eq.~2 with $H_c$ and $A$ as fitting parameters, while holding 
$\alpha = 0.48$.  Then the window was successively 
reduced by removing the highest temperature point, and the fitting procedure was 
repeated. The resulting $H_c$ values can then be plotted as a function of the upper 
limit of $T_W$, Fig.~5. The whole fitting procedure is then repeated for various 
fixed $\alpha$ values.  The value of $H_c = 7.2696 \pm 0.0002$~T was then obtained by 
taking a weighted average of the linearly extrapolated $H_c(T_W \rightarrow 0)$ 
values corresponding to various $\alpha$ values. This convergence of $H_c$ to a  
single value allows a more precise determination of $\alpha$ than possible with a 
three parameter fit in which $H_c$, $\alpha$, and $A$ are simultaneous fitting parameters. 
The critical exponent $\alpha$ is then determined by following the same procedure, where 
the $T$-window analysis is performed when fixing $H_c =7.2696$~T, Fig.~5, and the result is 
$\alpha = 0.481 \pm 0.004$.  
Establishing the critical exponent of the transition begs the questions about 
the details of the network of magnetic interactions, so 
inelastic neutron scattering experiments were performed.  

\begin{figure}[ht]
\includegraphics[width=3.375in]{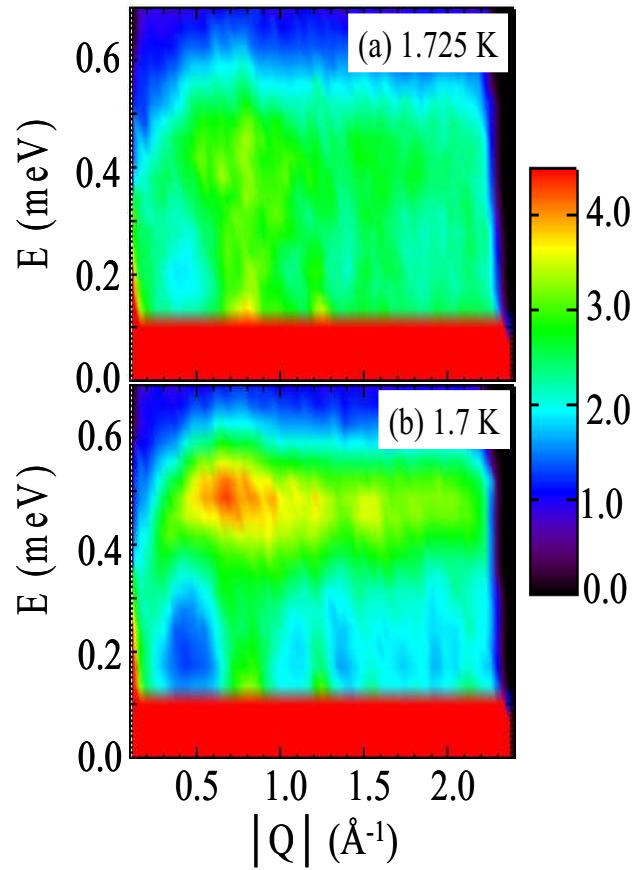}
\caption{(Color online) Inelastic neutron scattering spectra at (a) 1.725 K and (b) 1.7 K.}
\end{figure}

The neutron scattering pattern at 1.725~K (Fig.~6a) reveals a magnetic excitation 
above $\Delta E = 0.2$~meV. After decreasing the temperature to 1.7~K (Fig.~6b), this excitation 
becomes sharper with a well-defined mode near $\Delta E = 0.5$~meV. When the magnetic 
excitations are summed over all $Q$, the broad peak near $\Delta E = 0.5$~meV sharpens 
for $T < 1.725$~K. The temperature dependence of the integrated area of the peak is 
consistent with the LRAFO observed in the zero-field specific heat. 

\begin{figure}[ht]
\includegraphics[width=3.375in]{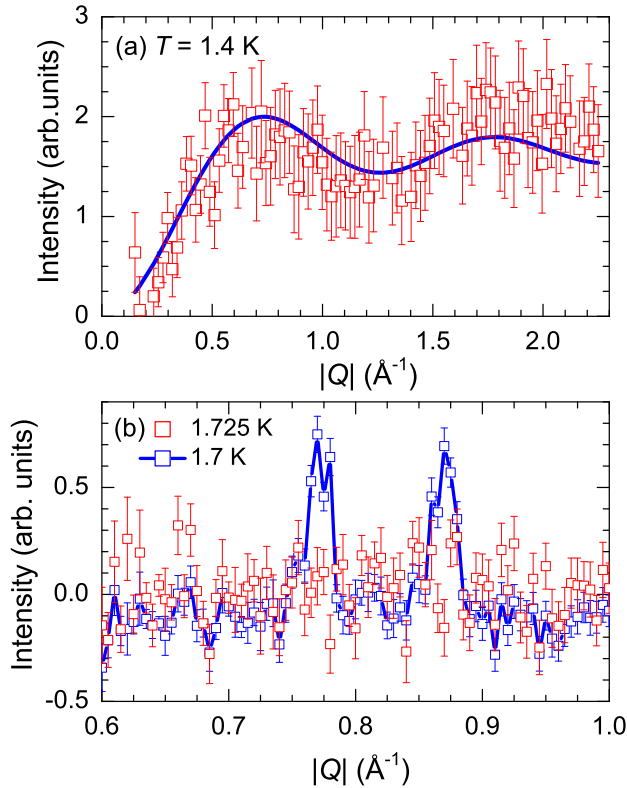}
\caption{(Color online) (a) A cut of the magnetic excitation near $\Delta E = 0.5$~meV at $T = 1.4$~K. 
The $Q$ dependence of the intensity (open squares) is fit to Eq.~3 (solid line). 
(b)~Appearance of the magnetic peaks at 1.7~K, after the data at 2~K 
have been subtracted as the lattice standard.}
\end{figure}

Another noteworthy feature is that the excitation below the transition shows clear $Q$ dependence, 
which has higher intensity at low $Q$. This magnetic contribution is consistent with an expression 
used to model the scattering from powders, where short-range, isotropic spin correlations 
persist to the first few coordination shells of neighboring spins,\cite{Bertaut} namely
\begin{equation}
I(Q) \;\sim\; \sum_{i,j}\,\langle \,s_i\,s_j\, \rangle\;\frac{\sin(Q\;r_{i,j})}{Q\;r_{i,j}}\;
\sim\;\frac{\sin(Q\;R)}{Q\;R}\;\;\;,
\end{equation}
where the final result is for the case of spins correlated over the nearest neighbors.  
Consequently, only one value $R$ of $r_{ij}$ (the distance between spins at sites $i$ and $j$) 
is employed in our fitting procedure, where Eq.~3, multiplied by the magnetic form factor 
of the Cr$^{5+}$ ion, has been used to fit to the net intensity shown in Fig.~7a.  This 
analysis yields $R = 6$~\AA, which is close to the $nn$ and $nnn$ Cr$\cdots$Cr 
distances that are in the range $5.8 - 6.2$~\AA. Ergo, the magnetic correlations are short range, 
extending over $nn$ and $nnn$ neighbors, Fig.~1. Furthermore, with decreasing temperature, 
the lattice Bragg peaks do not show obvious splitting or intensity change, 
thereby excluding the possibility of a structural change.  On the other hand, 
new magnetic Bragg peaks with low intensity develop below 1.725~K (Fig.~7b) and are consistent 
with the ordering of a small moment at $T_N \approx 1.7$~K.

Equipped with the result that short range magnetic interactions appear to extend over $nn$ and $nnn$ 
ions, the underlying network of these interactions can be reexamined.  
Previous theoretical and experimental studies established 
that although most of the unpaired electron density is on the Cr$^{5+}$ ion, a significant portion 
of it can be found on the bound (O$_2$)$^{2-}$ ions.\cite{McGarvey,Dalal1,Dalal2,Roch}  
Consequently, the potential superexchange pathways between Cr$^{5+}$ ions in K$_2$NaCrO$_8$ 
can be expected to involve the (O$_2$)$^{2-}$, Na$^+$ and K$^+$ ions. 
Since the bond angle is an important factor in determining the ferromagnetic versus antiferromagnetic 
nature of the interactions,\cite{Anderson,Goodenough,Hay} the paths with larger 
O$\cdots$Na/K$\cdots$O angle are selected when more than 
one possible exchange pathway for a given Cr$^{5+}$$\cdots$Cr$^{5+}$ distance are present.  
This analysis leads to the results given in Table I, which tabulates the 
most influential exchange interactions between $nn$ (the intra-planar $J_1$ and $J_2$)\cite{Choi} 
and $nnn$ (the inter-planar $J_3$ and $J_4$) Cr$^{5+}$ ions, Fig.~1. 
While the $J_1$ path involves 
Na$^+$ ions with a Na$\cdots$O distance of $\sim 2.4$~\AA~and a bond angle 
$\angle$ONaO of 142$^{\circ}$, the $J_2$ path 
involves K$^+$ ions with K$\cdots$O distance of $\sim 3.0$~\AA~and an angle 
$\angle$OKO of 166$^{\circ}$.   
The $J_3$ path involves Na$^+$ ions with a Na$\cdots$O distance of $\sim 2.4$~\AA~and 
a bond angle $\angle$ONaO of 166$^{\circ}$ and the $J_4$ path involves K$^+$ ions with a 
K$\cdots$O distance of $\sim 2.6$~\AA~ and a bond angle $\angle$OKO of 131$^{\circ}$.  
All of the data suggest the intra-layer 
interactions are somewhat stronger than the inter-layer ones, since the  
two-dimensional correlations develop before three-dimensional long-range order appears. 
Assuming a square lattice model, the intra-planar exchange interaction,  
$J_{\mathrm{intra}} \equiv (J_1 + J_2)/2$, can be estimated from the data.  For example, 
fitting $C_{\mathrm{mag}}(T, H = 18 \,\mathrm{T})$ yields $J_{\mathrm{intra}} \sim 2.1$~K, whereas 
$H_c = 4 J_{\mathrm{intra}}/g \mu_B$, gives $J_{\mathrm{intra}} \sim 2.4$~K. 
Both of these estimates are close to the value of 2.44~K obtained by fitting 
$\chi_{\mathrm{dc}}(T)$ using a two-dimensional high temperature series expansion.\cite{Cage}  

In summary, the low temperature, high magnetic field phase diagram of K$_2$NaCrO$_8$ 
has been established.  In the $T \rightarrow 0$ limit, the 
magnetic field tuned antiferromagnetic-ferromagnetic phase transition suggests 
a critical field $H_c = 7.270$~T and a critical exponent $\alpha = 0.481 \pm 0.004$. 
The value of this critical exponent, Eq.~2, depends on the universality class 
of the quantum phase transition, where $\alpha = 2/3$ is expected for Bose-Einstein 
condensation (BEC)\cite{Giamarchi1,Sachdev1,Sachdev2,Giamarchi2}  
and $\alpha = 1/2$ is predicted for Ising-like spins.\cite{Sachdev1,Sebastian2,Hamer}
The present data do not allow us to unambiguously differentiate 
between these two possible descriptions, since a broad range of exponents is expected if 
the transition is not probed at sufficiently low temperatures.\cite{Nohadani}  
The neutron data suggest the magnetic 
interactions extend over intra-planar nearest-neighbors and inter-planar 
next-nearest-neighbor spins.  For a deeper understanding of these results, 
especially the value of the critical exponent,   
inelastic neutron scattering experiments should be performed on single crystals 
that are cooled to millikelvin temperatures in the presence of the magnetic field 
capable of spanning $H_c$.  Fluctuation-induced heat release\cite{Kim} 
and NMR\cite{Suh} studies are alternatives that might provide decisive information. 

All experimental studies, except the ac susceptibility and neutron measurements, 
were performed at the National High Magnetic Field Laboratory (NHMFL), Tallahassee, 
which is supported by NSF Cooperative Agreement Grant No.~DMR-0654118 
and by the State of Florida. We thank C.~D.~Batista, K.~Ingersent, and D.~L.~Maslov for 
insightful conversations or communications, and we are grateful to Y.~Qiu and 
J.~R.~D.~Copley for help with the neutron measurements. This work was supported by NSF, in part, 
via DMR-0520481 (SN), DMR-0701400 (MWM), and DMR-0803516 (YL).

\begin{table}[hb]
\caption{Possible magnetic exchange pathways between Cr$^{5+}$ in K$_2$NaCrO$_8$.  
The labels for the atoms is the same as used by Cage and Dalal\cite{Cage} and the 
$J_1$ and $J_2$ designations are the same as used by Choi \emph{et al.}\cite{Choi}}
\centering
\begin{tabular}{c c c}
\hline
Label & Cr$\cdots$Cr (\AA) & Pathway\\
\hline
\hline
$J_1$ & 5.837 & O1$\cdots$Na2$\cdots$O2 \\
$J_2$ &  5.888 & O2$\cdots$K1$\cdots$O2 \\
$J_3$ & 5.974 & O6$\cdots$Na2$\cdots$O6 \\
$J_4$ & 6.243 & O3$\cdots$K1$\cdots$O3 \\
\end{tabular}
\end{table}


\end{document}